# Living Devices: The Physiological Point of View

Bob Eisenberg

*June 26, 2012*

Department of Molecular Biophysics and Physiology

Rush University

Chicago IL 60612

USA

## Abstract

The physiological tradition of biological research analyzes biological systems using reduced descriptions much as an engineer uses a 'black box' description of an amplifier. Simple models have been used by physiologists for a very long time. Physiologists have successfully analyzed a broad range of biological systems using a 'device-oriented' approach similar to the approach an engineer would use to investigate her devices. The present generation views biology through the powerful lenses of structural and (molecular) dynamic analysis, understandably enough because of the beauty and power of the analysis, and the ease of using these structures with present freely available software. The problem is that these powerful lenses offer such magnification that the engineering approach cannot be seen. High magnification means limited field of view, because the (spatial) dynamic range cannot cover everything. The function of the structures and molecular dynamics cannot be seen in the work of many biologists, probably because function cannot be immediately seen in the structures and molecular dynamics they compute. It is just as important for biologists to measure the inputs and outputs of their systems as it is to measure their structures. It seems clear, at least to one physiologist, that this research will be catalyzed by assuming that most biological systems are devices that can be analyzed with the same strategies one would use to analyze engineering devices. Thinking today about your biological preparation as a device tells you what experiments to do tomorrow. An important task for many of us is to transmit the physiological tradition to the next generation of biophysicists to help them adapt traditional questions to the new length scales and techniques of molecular and atomic biology.

The physiological tradition of biological research analyzes biological systems using reduced descriptions much as an engineer uses a 'black box' description of an amplifier.

An engineer is often not interested (to first order) in how the black box produces gain, or what inside the box produces gain, but studies the properties of the gain, its linearity, its frequency dependence and so on. She listens to the amplifier or uses it, almost the same way, whether it is made of tubes, bipolar transistors, or FETs (field effect transistors). A complete description of the structure of the amplifier or its internal physics is far less useful to the engineer than a reduced description of its input–output relation, when the goal is to use the amplifier or connect it to other devices to make a system.

An engineer told that an unknown black box is an amplifier is rather like a biologist confronting an ill understood biological system. Some structural knowledge is indispensable but complete structural knowledge is not. The engineer would have a terrible time if she did not know which wires were power supplies, which were inputs, and which were outputs. But the last thing the engineer would want to know is the physical layout of the amplifier. She would not even want a complete circuit diagram. A reduced circuit diagram of essential components—typically the input and output stages—would do fine. And she would have no idea what to do if she were given the atomic structure of the amplifier, i.e., the spatial coordinates of all molecules or atoms in its resistors, capacitors and transistors. The engineer wants to know enough to make the device work, or perhaps to improve or even construct the device, but no more. She has other things to do more useful than knowing everything that might be useful about the amplifier, let alone knowing the coordinate of every atom in one amplifier.

One of the most useful things the engineer can do is to measure the properties of the system, the input and outputs of the amplifier, under many conditions. Those inputs and outputs are related in a reproducible way, that can be described crudely with words, or more precisely by a device equation, or even better by a physical device model from which the device equation can be derived. Some structural knowledge is needed to interpret those measurements. Inputs, outputs, and power supplies have to be identified. Sometimes a complete circuit diagram is needed, but usually a simplified model is sufficient. Too much circuit or structural detail would be more a curse than a blessing. The complexity of a full description might obscure the essential components. Their functions would be hard to identify amidst all the less important detail.

Of course, the input output equation is not enough for all engineering purposes. Our argument is 'first things first'. First, understand the essential functions with a simple model. Then revise that model as needed. The simplest model of amplifiers are not general enough to describe the currents and voltages at the input and output as the devices connected to the amplifier are changed. If an amplifier is asked to measure tiny potentials, or currents, details of the input components and connections are very important, as can be glimpsed in papers that actually design amplifiers for biological use.[1-6] Situations like this depend on the sensitivity of the input voltage to the input current. Enough physical detail must be included to allow realistic computation of the sensitivity. But these considerations should follow the 'first order' analysis. Otherwise, the device may be seen as more complex than it really is. The simplest model should be sought first.

Simple models have been used by physiologists for a very long time. [7-11] Physiologists have successfully analyzed a broad range of biological systems using a 'device-oriented' approach similar to the approach an engineer would use to investigate an amplifier. For more than a century, medical students have used a device oriented approach to learn that the kidney filters blood to make urine; the lungs transport oxygen from air to blood; muscles contract; sodium channels produce action potentials; and so on. Each device description in physiology— on each length scale from organ, to tissue, to cell, to organelle (e.g., membrane), to protein molecule—is associated with a device model and equation, just as a device description in engineering (such as a sketch and verbal discussion of an amplifier or a solenoid) is followed by an approximate device model and equation for its input–output relation. Each device has an abstract representation in terms of its main function.

Each device also has a more complete description including the nonideal properties of its input and output, for example, the input and output impedances of an amplifier, and at a more detailed level yet, for the slew rate and nonlinear properties of the output of the (audio) amplifier. Each device has a still more detailed description designed to show how the outputs vary as conditions (e.g., power supply voltages) or temperature or input or output connections or components change. These more detailed descriptions involve the physics of some components inside the device and some of the connections of those components to each other. In none of these cases is a complete circuit description used, let alone a complete description of the layout (i.e., actual geometrical coordinates) of the components. No one has even thought of using a complete description of the location or movement of all the atoms of the device, as far as I know.

Until fairly recently, physiological analysis was focused on visible structures and systems and the device oriented approach was the only approach that could be used. The relation to the engineering approach was obvious to many 'old timers' but newcomers in the field, particularly those trained in the biochemical tradition, often did not notice the relation. Most of the models of physiology, particularly on the cellular and molecular length scale, were written in the language of chemical reactions, or reaction rate models that superficially appear similar to models of covalent bond change [12] but in reality are distinct [13-15] and the Appendix to [16]. But these rate models are in fact reduced models of a complex system in much the same spirit that engineering models are reduced. Molecular detail is suppressed whether the 'transfer function' is written in the language of electric circuits, or the 'binding function' or fluxes are written in the language of the chemical law of mass action.

Then came molecular biology and now even atomic biology (i.e., molecular biology in which the biological effects of individual atoms are significant and can be measured [17,18]). The immense work of scientists in structural biology and molecular dynamics now make it possible (even rather easy) to download structures of thousands of proteins in which the location of each atom is known. The movements of these thousands of atoms (in a single protein) can be computed with models of molecular mechanics that have been refined immensely since their (necessarily) primitive beginnings and now do remarkably well within limits, set by the atomic scale of size ($10^{-11}$ m) and motion ($10^{-16}$ sec). (The book by Tamar Schlick [19] provides a useful gateway into this immense field.) Obviously, it will be some time before molecular dynamics can reach the biological scale of time, size, and spatial resolution and compute the macroscopic functional variables that represent biological function [16]. Nonetheless, the immense resources devoted to this structural/dynamics approach has had profound success that has changed the entire approach of now two generations of scientists. The present generation views biology through the powerful lenses of structural and dynamic analysis, understandably enough because of the beauty and power of the analysis, and the ease of using these structures with present freely available software. The problem is that these powerful lenses offer such magnification that the engineering approach cannot be seen. High magnification means limited field of view, because the (spatial) dynamic range cannot cover everything. The function of the structures and molecular dynamics cannot be seen in the work of many biologists, probably because the function cannot be immediately seen in the structures and molecular dynamics they compute.

No one knows how much of biology can actually be computed without an explicit functional approach. No one knows which of the biological systems now being investigated on the molecular scale can be viewed productively as devices. No one knows how the magnificent structures of proteins so completely known on the atomic length scale determine the biological function of those proteins.

The contrast with the approach of physical scientists is striking. Think of the innumerable physicists and engineers interested in semiconductor devices for the last sixty years. The atoms of a semiconductor device are in a crystal lattice, far easier to compute and analyze than the disordered atoms of a protein or ionic solution. But physical scientists have not thought it necessary, nor I suspect feasible, to calculate the properties of the ordered atoms of their crystal lattices. Almost none have tried.

It is a wonderful testament to the enthusiasm and optimism of biologists that they are trying so hard to compute the motions of all the atoms of their proteins and the surrounding and interacting ionic solutions. One can only wish them luck, and remind them that when they enter realms where others fear to tread, comparisons with experimental measurements (on the experimental time and distance scales, in realistic biological conditions) are necessary to validate and calibrate their computations. Without validation and calibration, it is difficult to know what to make of the work.

It is necessary to show that the methods of molecular dynamics produce realistic descriptions of the properties of ions in water in the mixed solutions of $Na^+$, $K^+$, $Ca^{2+}$ and $Cl^-$ ions found inside and outside cells.[20] It is hard to believe that methods that cannot deal with divalents — or mixtures of salts like those found inside and outside cells — can actually compute the properties of a protein in those salts.

Handling this range of scales can make devices hard to simulate in atomic detail. Such detail may in fact be needed, when atomic details of structure are known experimentally to control biological function as is often the case when studying proteins or ion channels. Atomic time scale is rarely needed, but atomic spatial scales are needed in almost every case because atomic details of the structure of proteins are coded by genes to control specific macroscopic biological functions. Experiments of site directed mutagenesis show this in great detail. Individual mutations, that produce changes in structure on the atomic scale, directly control biological function on the macroscopic length and time scale. The challenge is to learn to compute on the atomic space scale and while averaging over some 10 orders of magnitude of time.

Devices can be difficult to investigate for other reasons. They can be complex and have interacting components and many internal nonlinear connections like the integrated circuit modules of digital computers or the central nervous system of animals. Systems can have overall function that only emerges when the entire system is connected. Such overall functions are not visible in the components, because they do not exist there. The overall functions depend on connections of components and particular properties of the components in a way that may become apparent only after all connections are made. Complex systems like this are hard to investigate because they are complex!

Complex systems may not be easily analyzed as devices, no matter how much experimental information is available, because they are complex, as described above. But one can also imagine simple systems—even as simple as an amplifier—that are hard to investigate only because of the paucity of experimental knowledge.

In fact, the largest problem facing the biologist has always been the lack of experimental knowledge.

If an engineer is given a black box, is told it is an amplifier, but is not told which wires are the power supply, input, or output; or if she is not told what are the specifications of the power supply (and not told which voltages damage the inputs), the investigation becomes more or less impossible. Reverse engineering of even simple systems is often 'ill posed'—mathspeak for "practically impossible"—simply because crucial information is missing. An entire branch of mathematics (called theory of inverse problems) has been developed to help squeeze useful information from ill-posed problems typical of reverse engineering. The math provides useful approximations to more of these ill posed problems than one would imagine. It is in fact possible to actually solve specific molecular problems, like the selectivity of models of ion channels, using the theory of inverse problems and its numerical methods.[21]

The experimental knowledge of proteins available routinely today would be (literally) unbelievable to my teachers. And one can only thank (and admire) our colleagues working in structural biology for such progress. This information provides us with indispensable knowledge of atomic function as well as a parts list of our devices. Both are obviously needed.

But this information remains a beginning, that must be supplemented by equivalent knowledge of function. If sufficient data about inputs and outputs of biological devices is not available along with the parts list and structural information, our inverse problems remain fundamentally (and probably impossibly) ill posed. If one applies the wrong voltages to a power supply of an amplifier, one gets behavior that is not relevant to its normal function. The normal function cannot be induced from the amplifier behavior until the power supplies are in the right range.

It is just as important for the next generation of biologists to learn to measure the inputs and outputs of the devices in their systems as it is to measure their structures, or develop theories of how the devices work when connected together.

It seems clear, at least to one physiologist, that this research will be catalyzed by assuming that most biological systems are devices that can be analyzed with the same strategies one would use to analyze engineering devices.

Thinking today about your biological preparation as a device tells you what experiments to do tomorrow. Thinking about preparations as devices helps design useful experiments Thinking about biological systems as devices will eventually lead to the device description, equation, and model, if they exist. Of course, not all biological preparations are devices, in the strict engineering sense of that word, just as all machines are not devices. Many machines do not have well defined inputs and outputs.

When investigating unknown biological systems, it is reasonable to first seek a description as a device. If no device description emerges after extensive investigation of a biological system, one can then look for other, more subtle descriptions of nature's machines. No one knows which of the unsolved complexities of biological research reflect problems of the reverse engineering of simple devices, and which reflect the inherent complexity of biological systems. But it is much more productive to assume that the device description exists than not. At least that way one knows what to do tomorrow!

> After all, many machines do not have well defined inputs and outputs, or even device descriptions. What is the input of a video game? Or the computer itself? What are the outputs? Useful abstract descriptions of machines like video games or computers are hard to construct, particularly if little is known about the machine and its use in the first place.

Physiologists have been thinking this way—specifically about their next experiments, or generally about life—for a long time, perhaps since Aristotle, and certainly before the development of engineering, molecular biology, or even biochemistry around the beginning of the $20^{th}$ century. It would be a setback if this device approach were lost to scientists of the $21^{st}$ century who have such extraordinary tools available that were unimaginable to their ancestors. But science needs questions as well as the tools to answer them. When the wrong outputs are sought, inverse problems are ill posed. When the wrong questions are asked, or no questions are asked at all, science proceeds slowly if at all. Answers to unasked questions are slow to emerge.

An important task for many of us is to transmit the physiological tradition to the next generation of biophysicists to help them adapt traditional questions to the new length scales and techniques of molecular and atomic biology.